\documentclass[aps,preprint,epsfig,multicol,nofootinbib,showpacs]{revtex4}
\usepackage{graphicx}
\usepackage{latexsym}

\begin{document}

\title{Dry Friction Avalanches: Experiment and Robin Hood model}

\author{Sergey V. Buldyrev$^1$, John Ferrante$^2$, and Fredy R. Zypman$^1$}
\affiliation{
$^1$Department of Physics, Yeshiva University
2495 Amsterdam Avenue, New York, NY 10033\\
$^2$NASA-Glenn Research Center, 
21000 Brookpark Road, Cleveland, OH 44135, USA\\}
\date{bfz.tex ~~~8~September~2005}

\begin{abstract}

This paper presents experimental evidence and theoretical models supporting
that dry friction stick-slip is described by self-organized criticality.  We
use the data, obtained with a pin-on-disc tribometer set to measure lateral
force to examine the variation of the friction force as a function of time.
We study nominally flat surfaces of aluminum and steel.  The probability
distribution of force jumps follows a power law with exponents $\mu$ in the
range 2.2 -- 5.4.  The frequency power spectrum follows a $1/{f^\alpha}$
pattern with $\alpha$ in the range 1 -- 2.6.  In addition, we present an
explanation of these power-laws observed in the dry friction experiments
based on the Robin Hood model of self organized criticality. We relate the
values of the exponents characterizing these power laws to the critical
exponents $D$ an $\nu$ of the Robin Hood model. Furthermore, we numerically
solve the equation of motion of a block pulled by a spring and show that at
certain spring constant values the motion is characterized by the same power
law spectrum as in experiments.  We propose a physical picture relating the
fluctuations of the force with the microscopic geometry of the surface.

\end{abstract}


\maketitle

\section{Introduction}

There are experimental and theoretical studies suggesting that certain far
from equilibrium systems with many degrees of freedom naturally organize in a
critical state, releasing energy through rapid relaxation events (avalanches)
of different sizes, these sizes being distributed according to a power law
probability density.  Examples of such behavior are found in earthquakes
\cite{Hallgass, Elmer}, biological systems \cite{Sole}, the stock market
\cite{Plerou}, rainfall \cite{Peters}, and friction
\cite{Slanina,Ciliberto,Vallette}. All these phenomena share the features of
the prototypical sand pile model \cite {Bak} for which the concepts of
self-organized criticality (SOC) were first proposed.  Recently, the
possibility of SOC \cite{Janson} in systems presenting stick-slip due to dry
friction has been under scrutiny \cite{Turcotte}.  In particular, Slanina
\cite{Slanina} presented theoretical attempts to explain dry friction in
terms of SOC. However, the central question remains unanswered: to what
extent is dry friction stick-slip a manifestation of SOC?  The clarification
of this issue has practical as well as fundamental implications.  From the
practical point of view, the power law exponents could be used as parameters
to characterize friction and wear of surfaces.  From the fundamental point of
view, there is a growing interest to understand systems driven far away from
equilibrium from a single unifying principle.  In addition, there is not yet
a full understanding of the dissipation mechanisms in friction. Particularly,
an overall description of the topography of the interface would be useful.
 
In the present study, we first present experimental results on stick slip in
dry friction using a pin-on-disc arrangement, set to measure lateral forces.
The probability distributions of force jumps sizes and the corresponding
frequency power spectra for aluminum and steel are examined for evidence of
SOC. Second, we present a theoretical explanation of the observed power laws
based on the Robin Hood model \cite{Zaitsev,Slanina}, which has been
successfully used in the past to study dislocation motion and friction.

\section{Experiment}

\begin{figure}[htb]
%\centerline
\includegraphics[width=12.0cm,height=12.0cm,angle=0]{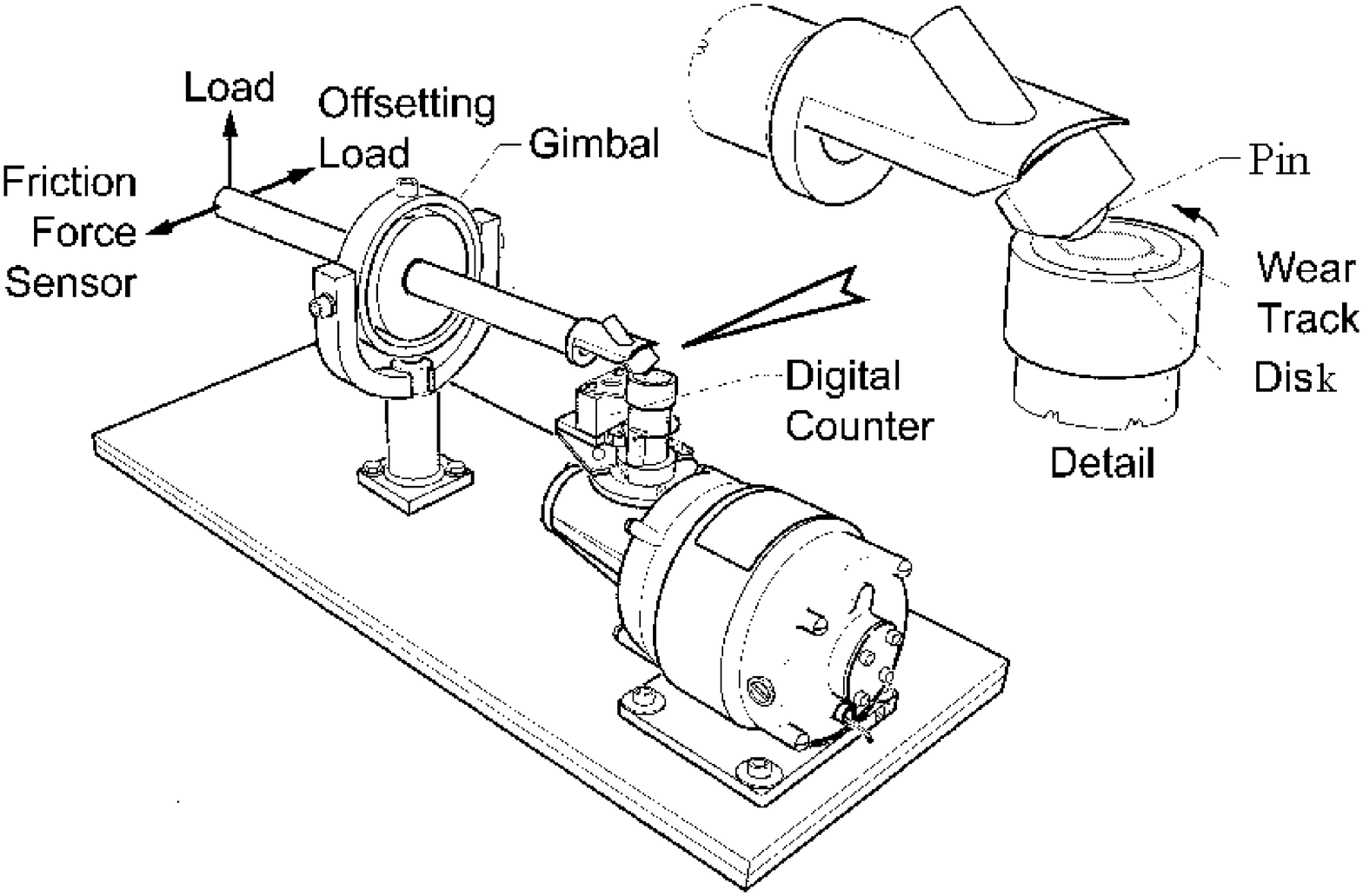}
%}
\caption{Pin-on-disk tribometer.  The arrow points at the 
location where the spherical pin and the disk touch.  The disk lies 
horizontally while the pin attached to the arm, rests above it.
\label{X1}}
\end{figure}

\begin{figure}[htb]
%\centerline
\includegraphics[width=12.0cm,height=12.0cm,angle=0]{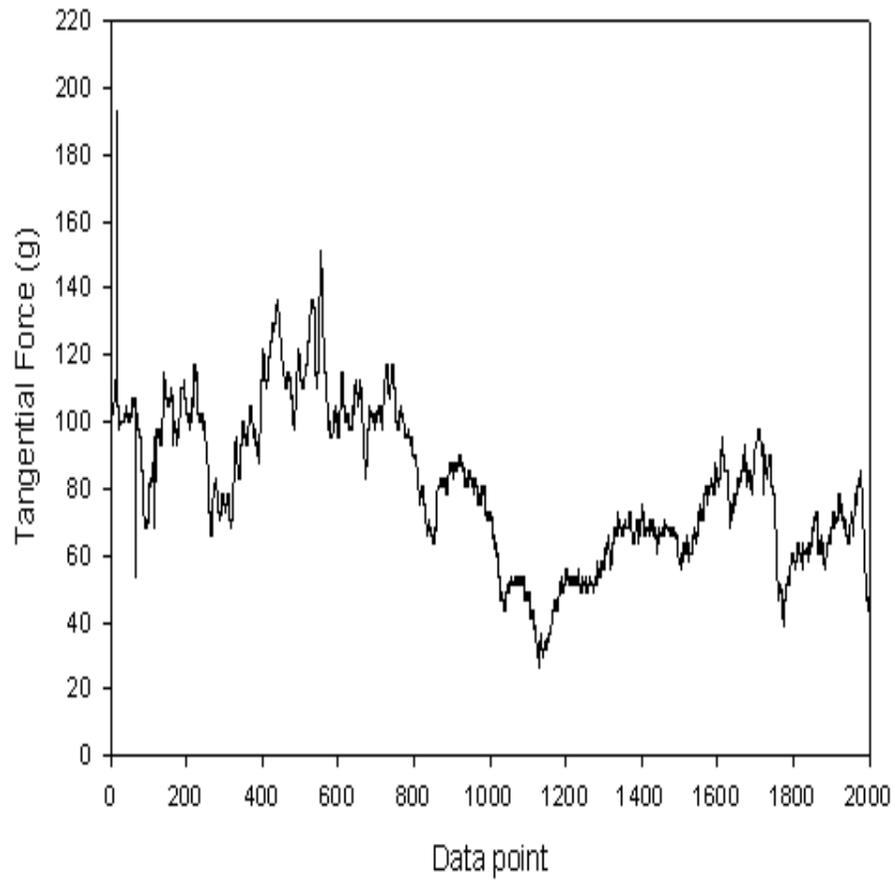}
%}
\caption{Typical signal from the tribometer.  The effective 
spring constant of the apparatus is $~1$g/${\mu}$m, giving the 
largest force jumps as a few hundred ${\mu}m$.
\label{X2}}
\end{figure}

The pin-on-disk tribometer used for these experiments is
shown in Fig.~\ref{X1}.  This configuration was chosen because it
allows for easy replacement of the contacting surfaces, and it is the
standard method for measuring friction and wear in unlubricated and
lubricated contacts.  The apparatus uses a $2.54$ cm diameter disk and a
spherical pin with a $0.95$ cm radius machined on its end.  The pin is attached
to a load arm that is mounted on a gimbal supported at the center
through which a load applied at the end of the arm is transferred to
the contact zone.  A strain-gauge is mounted at the end of the arm to
monitor tangential friction force.  The tangential force is monitored
at a sample rate of 1,000 scans/sec and conversion is done using a
16-bit data acquisition card controlled by LabVIEW.  Data is recorded to a text
file for later analysis.  Frictional force measurements are done on
matching aluminum and steel (M50) pin-and-disc tribometers.  The
signal is collected at 1KHz during 16min, thus collecting $10^6$ points.
The first quarter of each data set, or about 4 minutes, is
discarded to assure that steady state is reached.
In order to drive the system very slowly away from static equilibrium,
we select slow rotational speeds in the range 10-20 RPM.  
Each disk is used for up to
four tests by changing the radial position of the pin on the disk.
Loads for aluminum range from 250 g to 1000 g. Steel is studied with a 1000 g normal load
between pin and disk.    Figure~\ref{X2} shows a
typical tangential friction force time series.  It shows force jumps of
various sizes.  We first construct the
probability distribution of force jumps. Force jumps are taken as those events 
corresponding to negative changes in the tangential force.   

\begin{figure}[htb]
%\centerline
\includegraphics[width=12.0cm,height=12.0cm,angle=0]{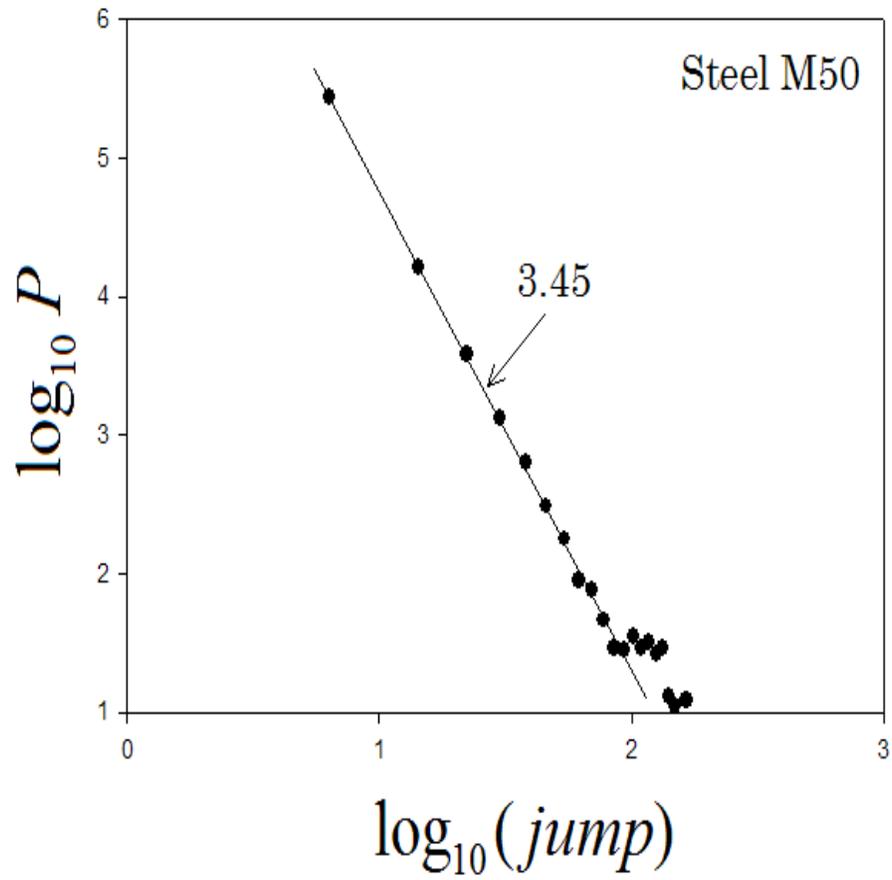}
%}
\caption{Probability density for jump size distribution on
steel M50.
\label{X3}}
\end{figure}

\begin{figure}[htb]
%\centerline
\includegraphics[width=12.0cm,height=12.0cm,angle=0]{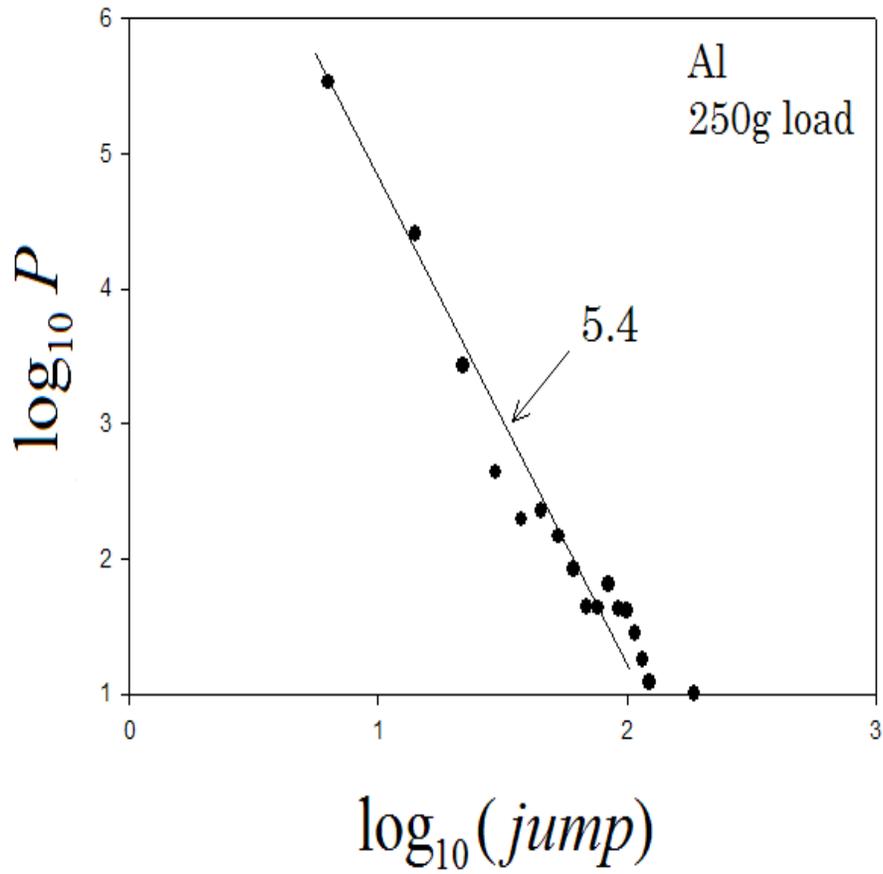}
%}
\caption{Probability density for jump size distribution
on aluminum with a normal load of 250g.
\label{X4}}
\end{figure}

\begin{figure}[htb]
%\centerline
\includegraphics[width=12.0cm,height=12.0cm,angle=0]{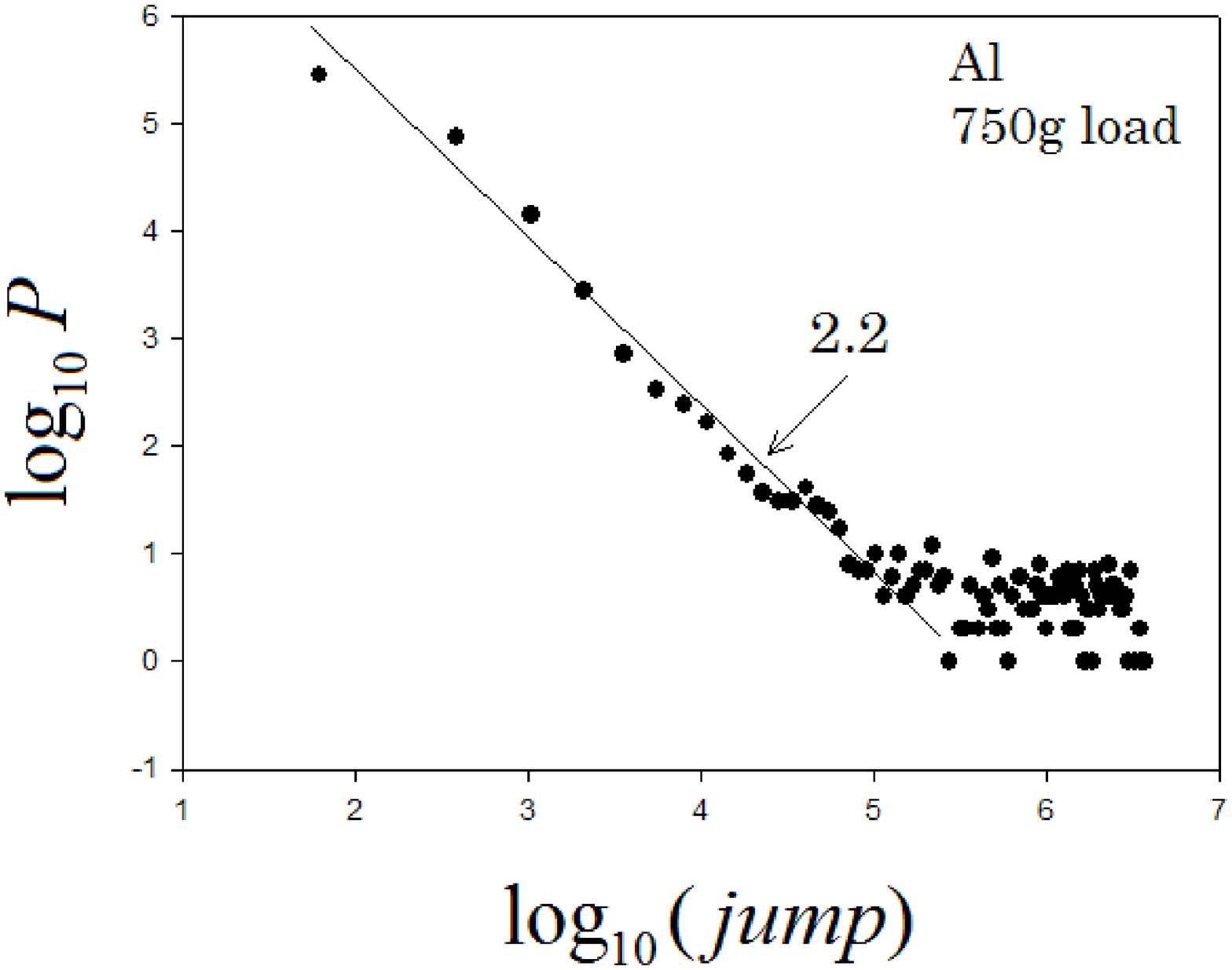}
%}
\caption{As in Fig.~\ref{X4} with normal load of 750g.
\label{X5}}
\end{figure}

\begin{figure}[htb]
%\centerline
\includegraphics[width=12.0cm,height=12.0cm,angle=0]{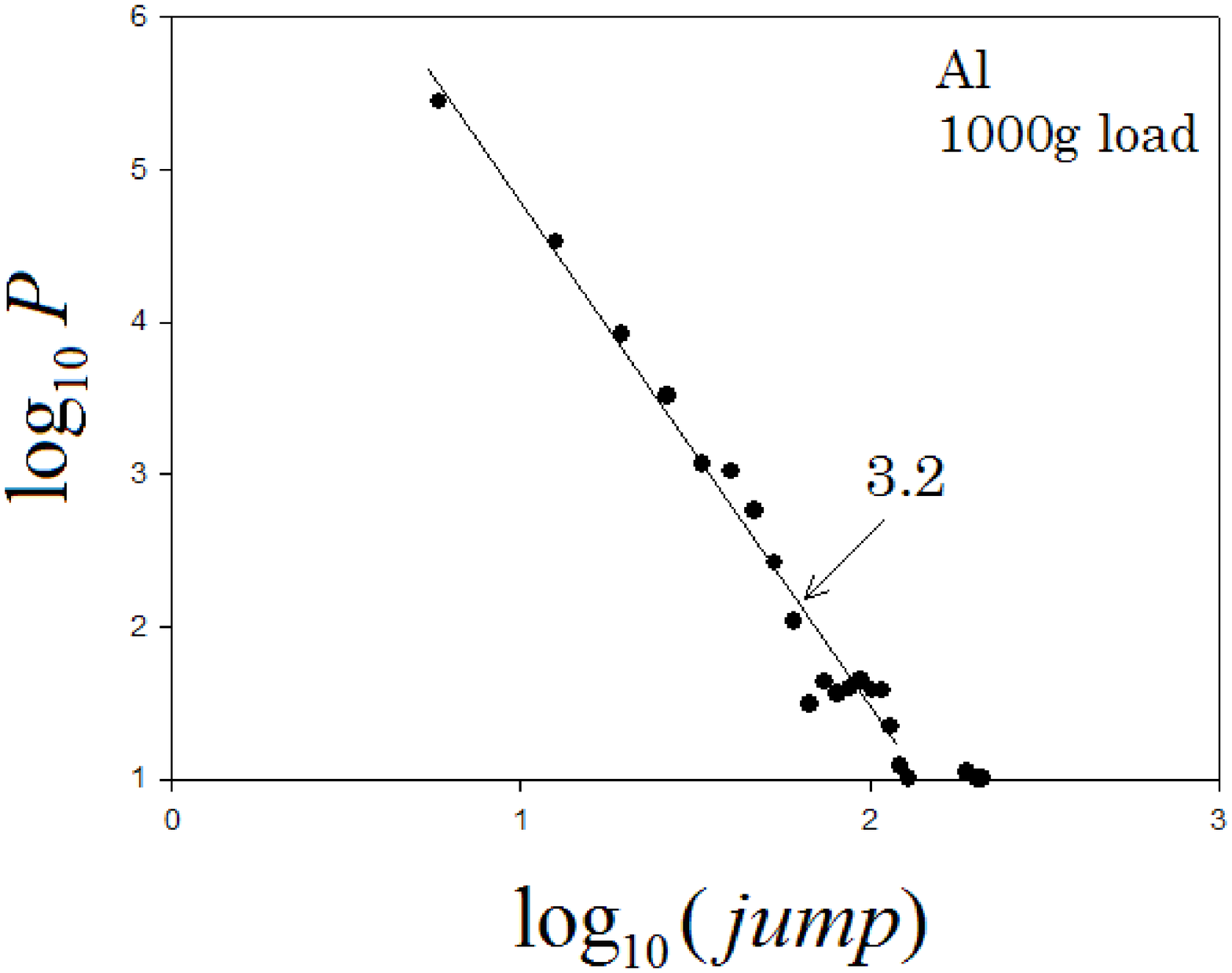}
%}
\caption{As in Fig.~\ref{X4} with normal load of 1000g.
\label{X6}}
\end{figure}

We next obtain the probability distributions of the tangential friction 
forces corresponding to aluminum under various loads and M50 steel.
Results of such analysis are presented in Figs. \ref{X3} through
\ref{X6}. We observe an approximate linear behavior on
the double logarithmic plots suggesting the power law behavior of the 
distributions $P(F)\sim F^{\mu}$ with  the exponents $\mu$ in the range
between 2.2 and 5.4.

\begin{figure}[htb]
%\centerline
\includegraphics[width=12.0cm,height=12.0cm,angle=0]{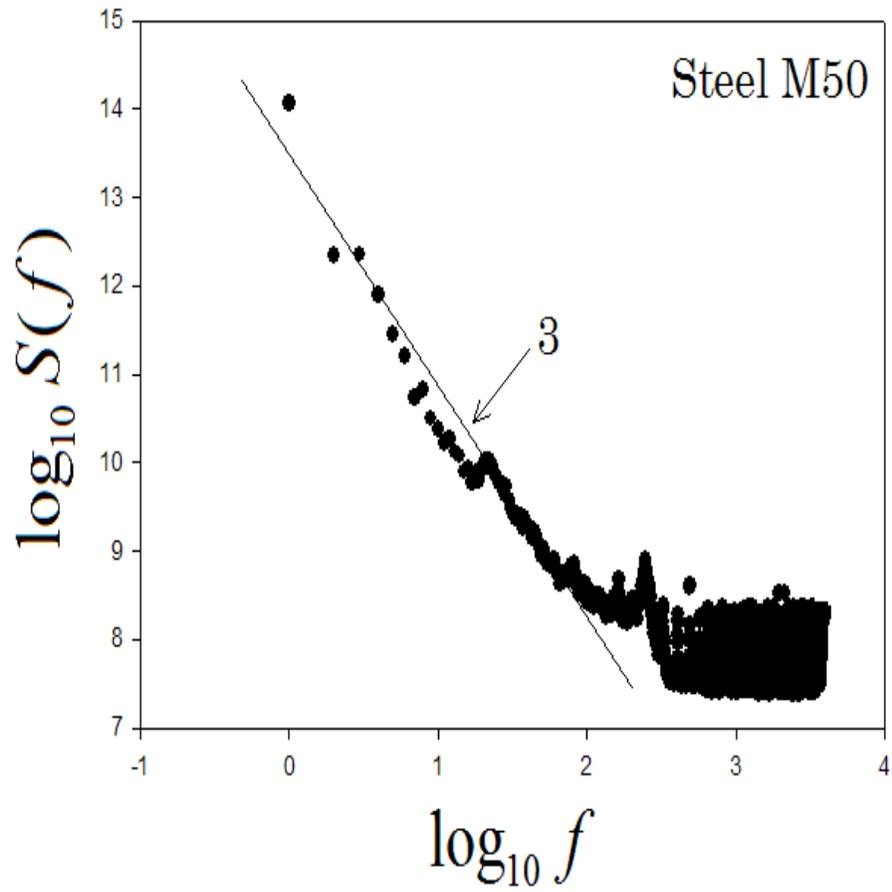}
%}
\caption{Power spectrum of jump sizes for a
steel M50 sample.
\label{X7}}
\end{figure}

\begin{figure}[htb]
%\centerline
\includegraphics[width=12.0cm,height=12.0cm,angle=0]{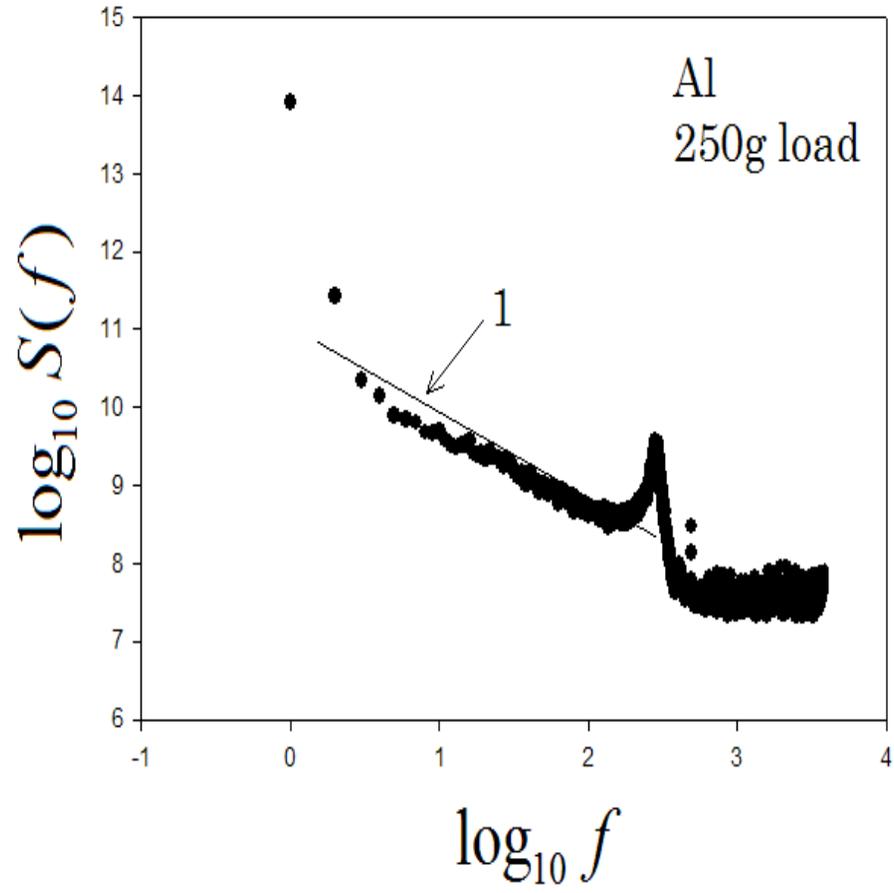}
%}
\caption{Power spectrum of jump sizes for
an aluminum sample with a normal load of 250g.
\label{X8}}
\end{figure}

\begin{figure}[htb]
%\centerline
\includegraphics[width=12.0cm,height=12.0cm,angle=0]{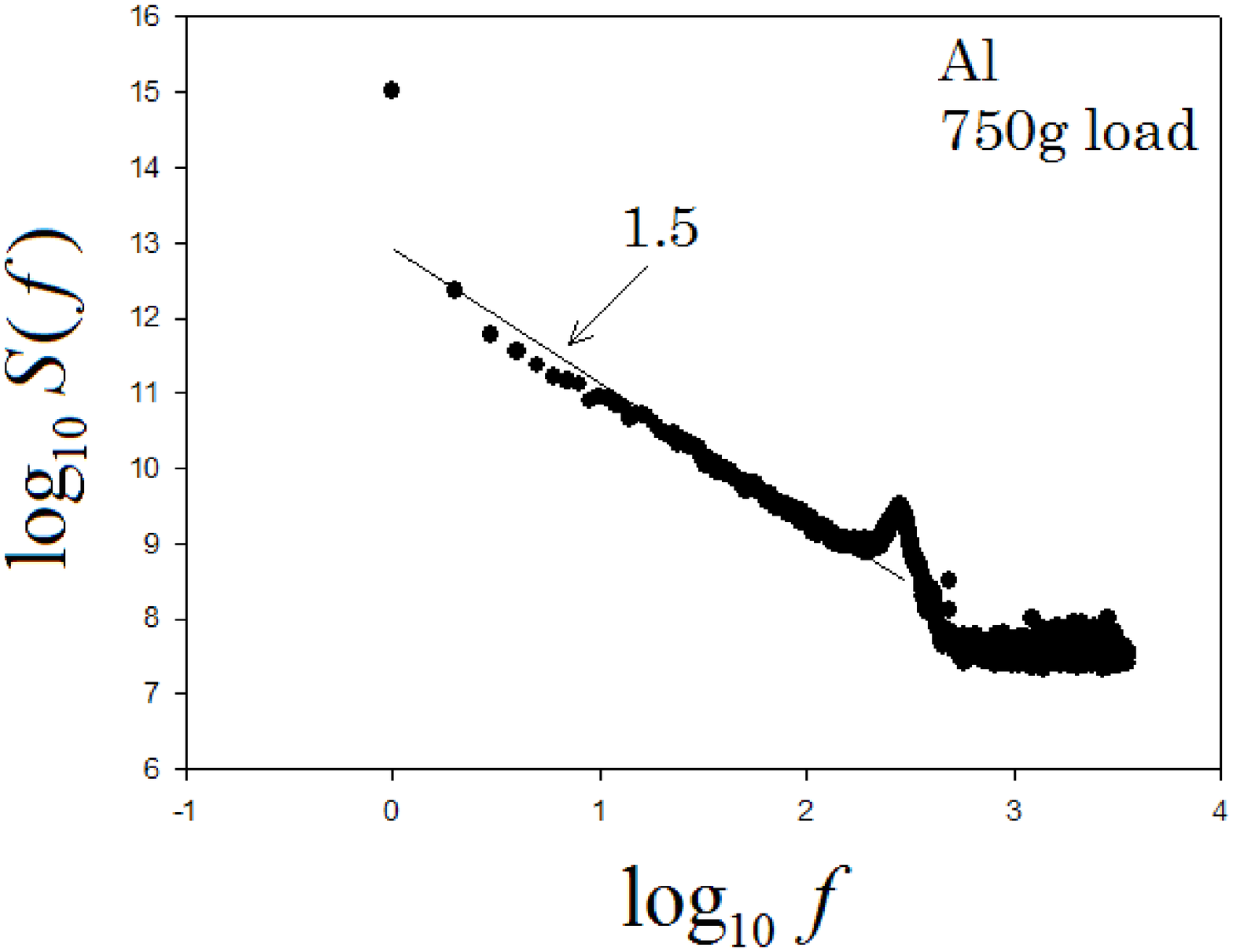}
%}
\caption{As in Fig.~\ref{X8} with a normal load of 750g.
\label{X9}}
\end{figure}

\begin{figure}[htb]
%\centerline
\includegraphics[width=12.0cm,height=12.0cm,angle=0]{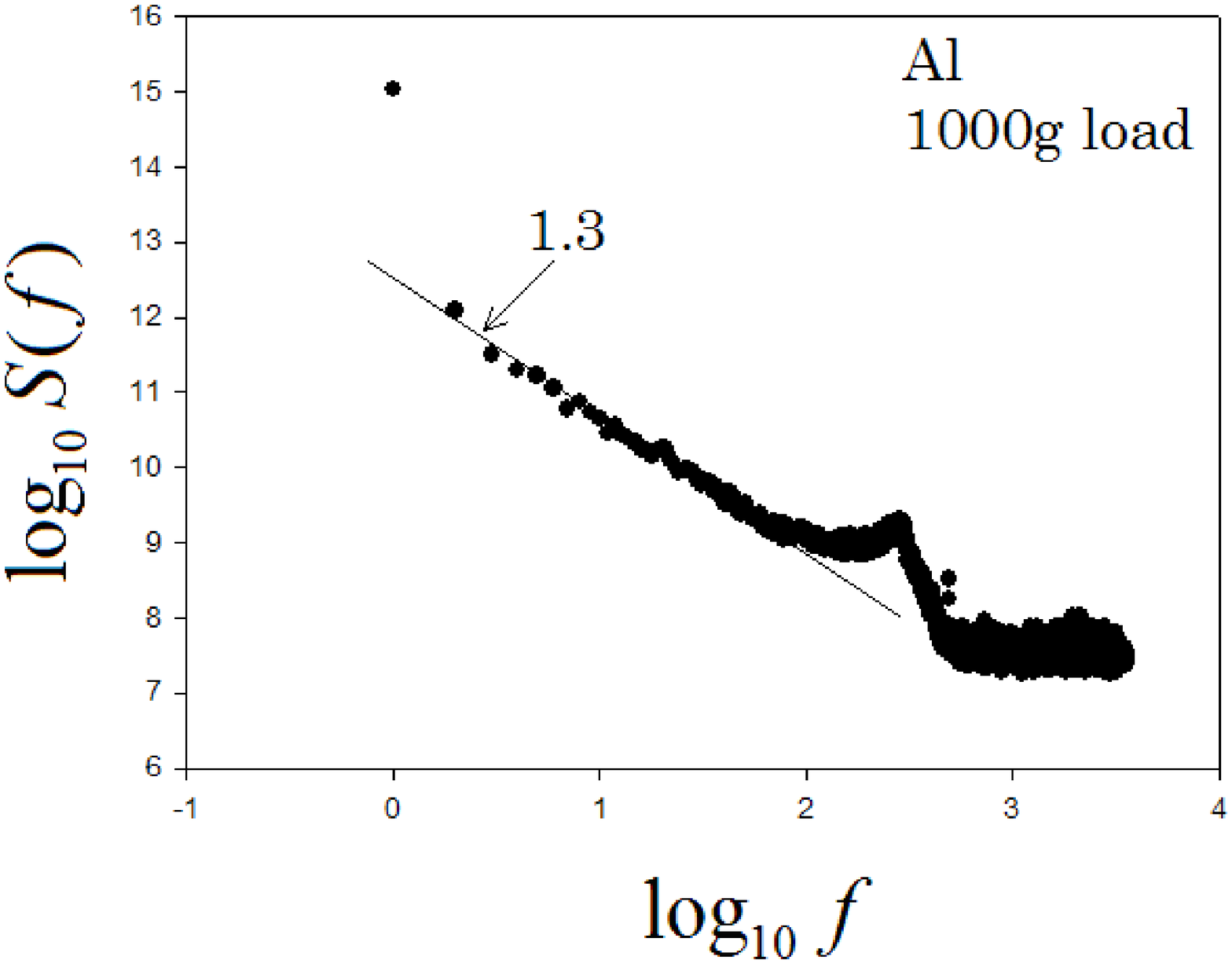}
%}
\caption{As in Fig.~\ref{X8} with a normal load of 1000g.
\label{X10}}
\end{figure}

We also compute the power spectra of the tangential friction force time
series.  We divide the original data ($2^{19}$) into $2^8$ statistically
independent non-overlapping data sets of $2^{11}$ points each. Next, we
calculate $2^8$ individual power spectra and finally average them to obtain
the resulting power spectrum. Specific double-logarithmic plots are shown in
Figs.~\ref{X7}-\ref{X10}.  The power spectra follow power laws with
exponents  $1.0\leq\alpha \leq 2.6$. The results of the probability
distribution and power spectra analysis are summarized in Table I.

\section{Theory}

Friction is believed to occur on the atomic scale due to asperities on the
surfaces in contacts. The simplest way to model such asperities is to use
lattice models in the spirit of the invasion percolation \cite{Guyon}, the
sandpile model \cite{Bak}, Bak-Sneppen evolution model \cite{Sneppen}, and
Zaitsev's Robin Hood model \cite{Zaitsev}. In general, these models are too
crude to provide quantitative agreement with all experimental statistical
quantities. However, quantities such as distribution of avalanche sizes are
usually described by power laws characterized by exponents belonging to a few
distinct universality classes. These exponents may be compared with
experimentally found ones. If several models belong to the same universality
class, it is reasonable to study the simplest among them, since it will
provide the clearest understanding of the physical mechanisms of the
phenomenon under study. One such simple and elegant example is the Robin Hood
model, which was originally proposed for dislocation movement \cite{Zaitsev}
and later was adopted for modeling dry friction \cite{Slanina} by Slanina who
added to the original Robin-Hood model several parameters aimed to better
capture the dry friction mechanism, but essentially obtained the same type of
behavior as the original model had.  Here we return to the original Robin
Hood model due to its simplicity.

The model consist of a $d$-dimensional lattice.  Each site $i$ on this
lattice at any time step $n$ is characterized by the height $h_i(n)$
which we assume to be the height of an atomic scale asperity at a given point
of the interface between two bodies in contact. Here we present the model for
$d=1$, which is an appropriate choice to model slide friction but the
analytical treatment is the same in any dimension, although the physically
relevant cases are only $d=1$ and $d=2$. As the bodies slide against each
other, the asperity with the maximal height is destroyed and some random
number of atoms from this asperity is distributed among the neighboring
asperities. To be specific, at each time step the site $i$ with maximal
height $h_m(n)=\max h_i(n)$ is found and the new heights are determined
according to the following rule: $h_m(n+1)=h_m(n)-r(n)$ and $h_{m\pm1}(n+1)=
h_{m\pm1}(n)+r(n)/2$, where $r(n)$ are independent random variables uniformly
distributed between 0 and 1.  (Robin Hood determines the richest merchant in
the market, robs him by a random amount $r(n)$ and distributes it equally
among the neighbors without leaving anything for himself). If we assume
periodic boundary conditions so that the sites with $i=0$ and $i=L$ are
equivalent, the total amount of matter $\sum_{i=0}^L h_i(n)$ is conserved and
we can assume it to be zero.  The distance between the surfaces at a given
site $i$ can be determined as $h_m(n)-h_i(n)$. 

The particular details of the model such as the
distribution of $r(n)$ or the rule of dividing it among neighbors can vary, but
the model still retains its SOC behavior . The critical exponents appear to be
sensitive to the details of dividing  $r(n)$, for example, an exactly
solvable asymmetric model (in which all the profit is given to the site on
the left) \cite{Maslov95} belongs to a different universality class.

It has been shown\cite{Maslov,Paczuski} that in a wide class of
depinning SOC models, all the critical exponents can be expressed in
terms of the two main exponents : avalanche dimension $D$ and
correlation exponent $\nu$.  The avalanche of threshold $h_0$ is
defined as sequence of time steps during which the hight of the
maximal asperity is above $h_0$. Namely, if $h_m(n_0)\geq h_0$ and
$h_m(n_0+s)\geq h_0$ while for $n_0<n<n_0+s$, $h_m(n)<h_0$, the sequence
$n=n_0+1,.., n_0+s$ is called a punctuating avalanche of threshold
$h_0$ and mass $s$. The avalanche dimension
describes how the avalanche mass $s$, scales with the horizontal
dimension of the avalanche $R$. To be more precise, the mass
distribution of forward avalanches with threshold $h_0$, scales as
\begin{equation}
P_s(s)\sim s^{-\tau_s}g_s(s(h_0-h_c)^{D\nu})
\label{P_s}
\end{equation}
and the distribution of the avalanche horizontal size, $R$, scales as
\begin{equation}
P_R(R)\sim R^{-\tau_R}g_R(R(h_0-h_c)^{\nu}),
\label{P_R}
\end{equation}
where $h_c \approx 0.114$ is the critical height,
\begin{equation}
\tau_s=1+(d-1/\nu)/D 
\label{tau_s}
\end{equation}
and 
\begin{equation}
\tau_R=1+d-1/\nu
\label{tau_R}
\end{equation}
are Fisher exponents first introduced to characterized cluster distributions in
percolation theory \cite{Stauffer}, while
$g_s$ and $g_R$ are exponentially decreasing cutoff functions.
It has been suggested \cite{Roux}, that the Robin Hood model belongs
to the same universality class as the linear interface model, for which
the values ($D=2.23$ and $\tau_s=1.13$ in $d=1$; $D=2.725$ and $\tau_s=1.29$
in $d=2$) are given in Ref.\cite{Paczuski}. Using these values 
and Eq. (\ref{tau_s}), one gets $\nu=1.41$ for $d=1$ and $\nu=0.83$ for $d=2$. 

It can be shown that after the initial equilibration number of time 
steps $T\sim L^D$,
any initial shape of the interface $h_i(0)$ reaches a steady state
such that very few $N(L)$ ``rich'' sites have $h_i(n)>h_c\approx 0.114$, where
\begin{equation}
N(L)\sim L^{d_f} 
\label{N_L}
\end{equation}
and 
\begin{equation}
d_f=d-1/\nu
\label{d_f}
\end{equation}
plays the role of fractal dimension of rich sites.  Only those few
rich sites have a chance to be robbed. The chance $P_m(h_m)$ that at a
given  time step, the maximal height is equal to $h_m$ decreases
for an infinite system \cite{Maslov} as 
\begin{equation}
P_m(h_m)=(h_m-h_c)^{\gamma-1}, 
\label{P_r}
\end{equation}
where the exponent 
\begin{equation}
\gamma=1+\nu(D-d)
\label{gamma}
\end{equation}
characterizes the dependence of the average avalanche size on its
threshold $h_0$: 
$\langle s\rangle\sim (h_0-h_c)^{-\gamma}$.  

The distribution of 
heights of the poor sites converges to a smooth distribution on
the interval $[h_c-1,h_c]$, while the distribution of the rich sites
converges to the distribution with a power law singularity 
\begin{equation}
P_h(h)\sim(h-h_c)^{-d\nu}.
\label{P_h}
\end{equation}
This result is not presented in Refs.
\cite{Maslov,Paczuski} but can be justified by the following
heuristic arguments. Indeed, the number of sites with $h>h_0$ scales as
the number of the active sites in an avalanche of threshold $h_0$,
and thus scales as $R^{d_f}(h_0)$, where $R(h_0)$ is the cutoff
of the avalanche distribution (\ref{P_R}) which scales as 
\begin{equation}
R(h_0)\sim(h_0-h_c)^{-\nu}.  
\label{R_h_0}
\end{equation}
Thus the probability that $h>h_0$ scales as
$(h_0-h_c)^{-d_f\nu}$ 
and the probability density of $h=h_0$ scales as
\begin{equation}
P_h(h_0)\sim(h_0-h_c)^{-d_f\nu-1}= (h_0-h_c)^{-d\nu}. 
\label{P_h1}
\end{equation}

We can assume that the
friction force, $F(n)$, at a given time step is proportional to the
number $P_h(h_m(n))\Delta h$ of asperities with heights between 
$h_m(n)-\Delta h$ and $h_m(n)$: 
\begin{equation}
F(n)= F_1 P_h(h_m(n))\Delta h, 
\label{F_h}
\end{equation}
where $\Delta h$ is the interaction distance of atomic forces acting between
the two surfaces and $F_1$ is a proportionality coefficient, corresponding
to the surfaces interaction force at the asperity. Accordingly, the
distribution of the friction forces $P(F)$ satisfies the equation $P(F)d
F=P_m(h_m)d h_m$, where the random variables $F$ and $h_m$ are linked by
Eq. (\ref{F_h}). Taking into account Eqs. (\ref{P_h}) and (\ref{F_h}) we have
$dh_m/dF\sim d F^{-1/d\nu}/dF \sim F^{-1/d\nu -1}$. Finally Eq. (\ref{P_r})
yields
\begin{equation}
P(F)=
P_m(h_m(F)){d h_m\over d F}\sim F^{-(\gamma-1)/d\nu}
F^{-1/d\nu-1}=F^{-\mu},
\label{P_F}
\end{equation}
where 
\begin{equation}
\mu=(D+1/\nu)/d.
\label{mu}
\end{equation}
For $d=1$, using
values of Ref.\cite{Paczuski} we have $P(F)=F^{\mu}$ with $\mu=2.94$ which is
consistent with the experimental observations of the density of jump
sizes presented here and in Ref.\cite{Zypman}. For $d=2$ we have $\mu=1.96$.

\begin{figure}[htb]
\includegraphics[width=12.0cm,height=12.0cm,angle=270]{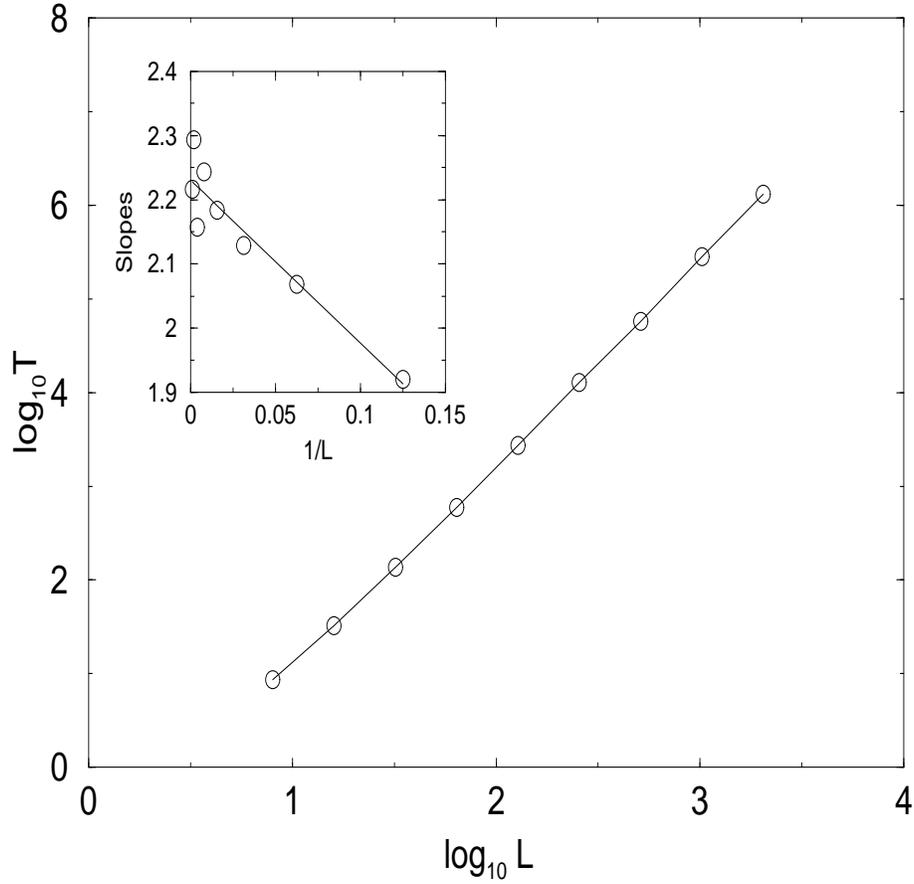}
\caption{ Double logarithmic plot of the average equilibration time
 $\langle T\rangle$  versus system size $L=2^3,2^4, ..., 2^{12}$.
 The inset shows the successive slopes of the main graph versus $1/L$.
 The intercept $D=2.23$, agrees with the data of Ref.\cite{Paczuski}.
 \label{T-L}}
\end{figure}

In order to test this theoretical predictions, we perform
simulations of the one dimensional Robin Hood model. Starting at $n=0$ with
a flat interface $h_i(0)=0$, and selecting the first site to rob at
random, after $T$ steps we get all $L$ sites of the interface updated
at least once. Measuring the average $\langle T\rangle$ for many
independent runs for different system sizes, and plotting it versus $L$
in a double-logarithmic scale (Fig. \ref{T-L}), we can obtain the
avalanche dimension $D$ as the limit of the successive slopes of this
graph for $L\to\infty$.

\begin{figure}[htb]
\includegraphics[width=12.0cm,height=12.0cm,angle=270]{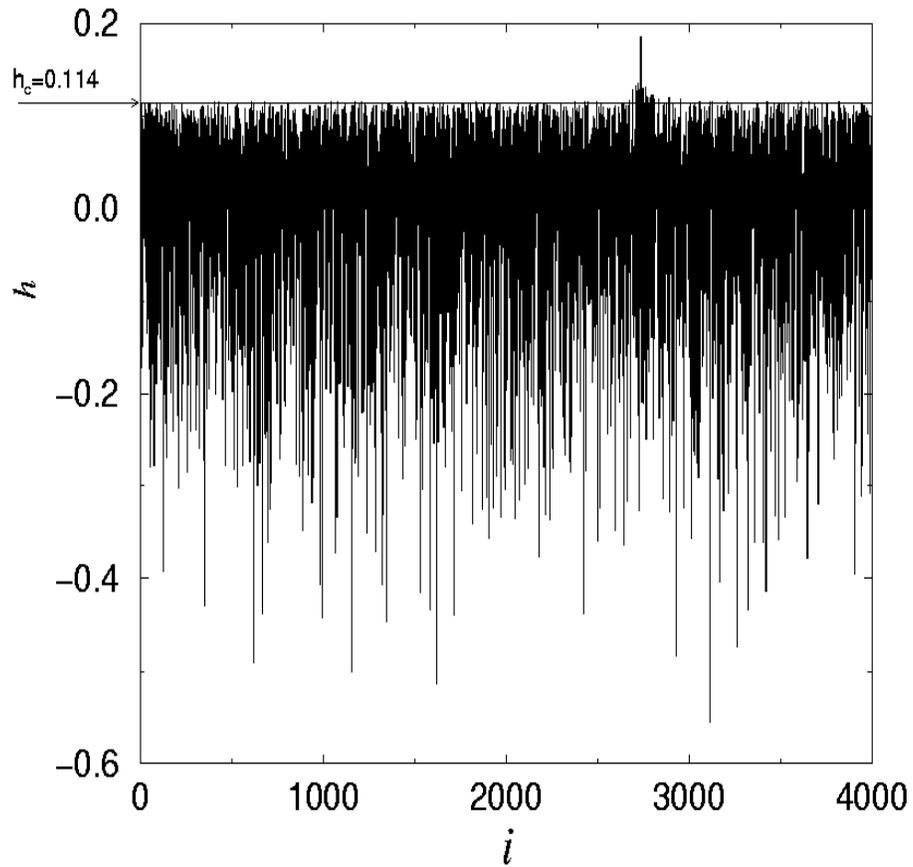}
\caption{ A typical shape of the steady state interface. The horizontal
line shows the critical height.   
\label{h-i}}
\end{figure}

A typical shape of the interface at time $n>T$ is presented in
Fig. \ref{h-i}.  One can see that the height of the majority of sites
do not exceed the critical value $h_c\approx 0.114$. Interestingly,
the majority of rich sites with heights above the critical barrier are
localized in the vicinity of the richest site.

\begin{figure}[htb]
\includegraphics[height=9.0cm,angle=270]{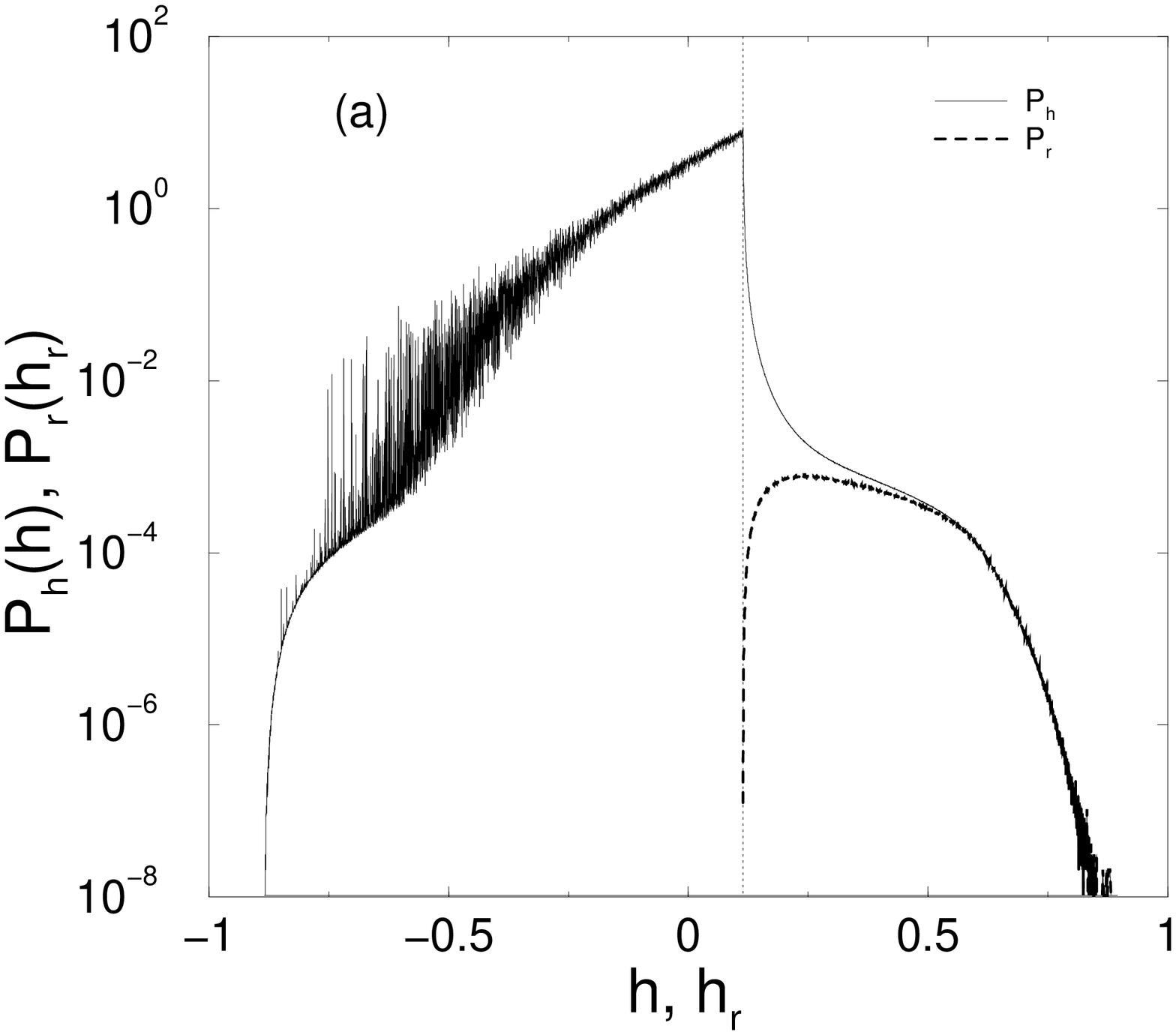}
\includegraphics[height=9.0cm,angle=270]{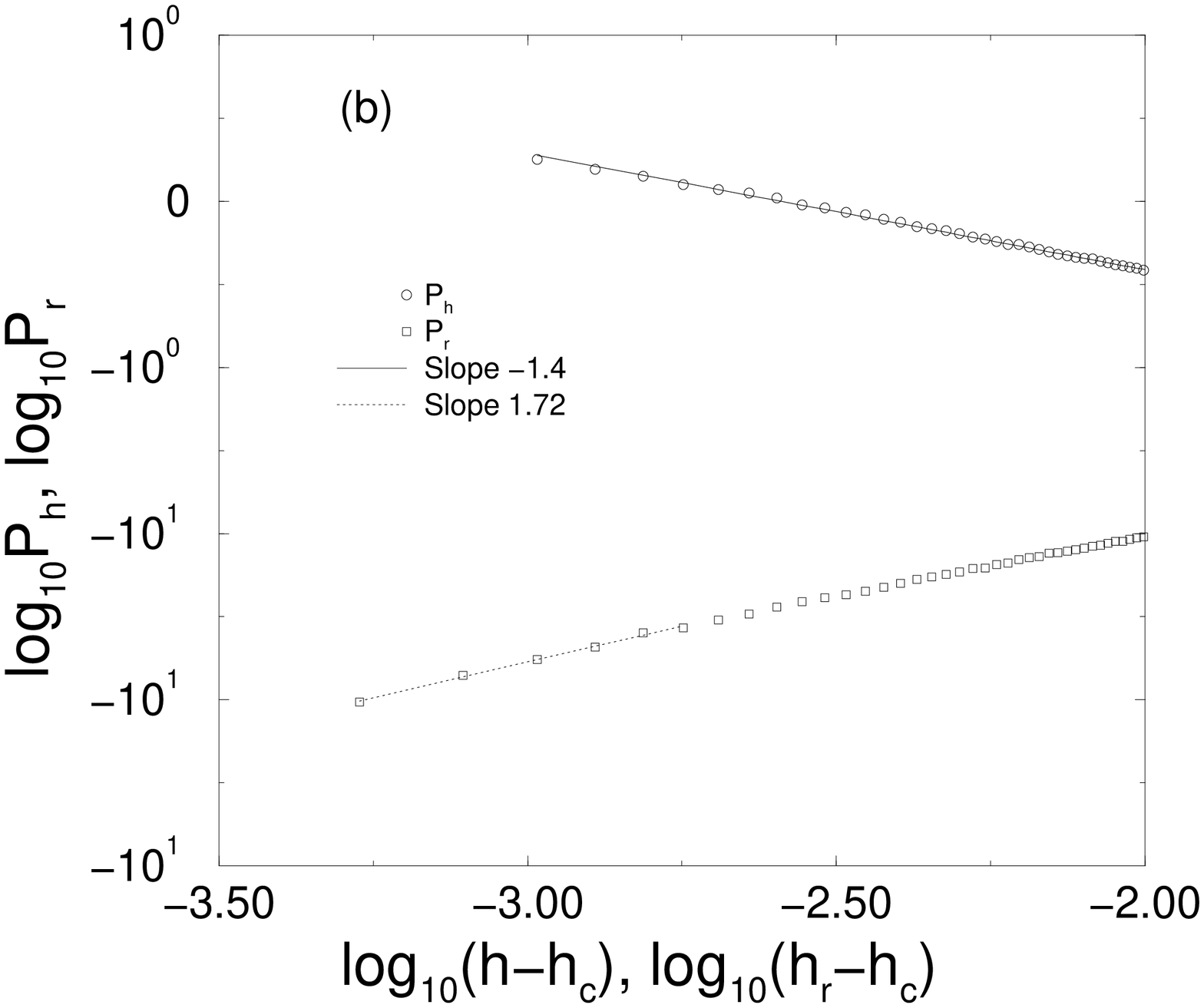}
\caption{ (a) The semi-logarithmic plot of the distribution of heights
$P_h(h)$ of all sites (solid line) and the distribution of heights
$P_m(h_m)$ of robbed sites (dashed bold line).
Vertical dotted line shows the position of the critical height $h_c=0.114$.
(b) Double logarithmic plot of the same quantities plotted as functions of
$h-h_c$. The slopes of the curves in the fitted regions are
in agreement with Eq.(\ref{P_h}) $-d\nu=-1.4$ and Eq. (\ref{P_r}) $\nu(D-1)=1.72$.
\label{hist}}
\end{figure}

Figure \ref{hist}
shows the histogram of all the interface heights $P_h(h)$ collected over 
many time steps after the system has reached the steady state and the
histogram of the heights of the robbed sites $P_m(h_m)$. One can see that
while  $P_h(h)$ dramatically increases as $h\to h_c^+$, no sites below
the critical value are robbed and $P_m(h_m)\to 0$ as $h_m\to h_c^+$.
In order to find the exponents governing the behavior of these distributions
near the critical point, we plot these quantities in a double logarithmic 
scale as functions of $h-h_c$ (Fig.~\ref{hist}b).

\begin{figure}[htb]
%\centerline
\includegraphics[width=12.0cm,height=12.0cm,angle=270]{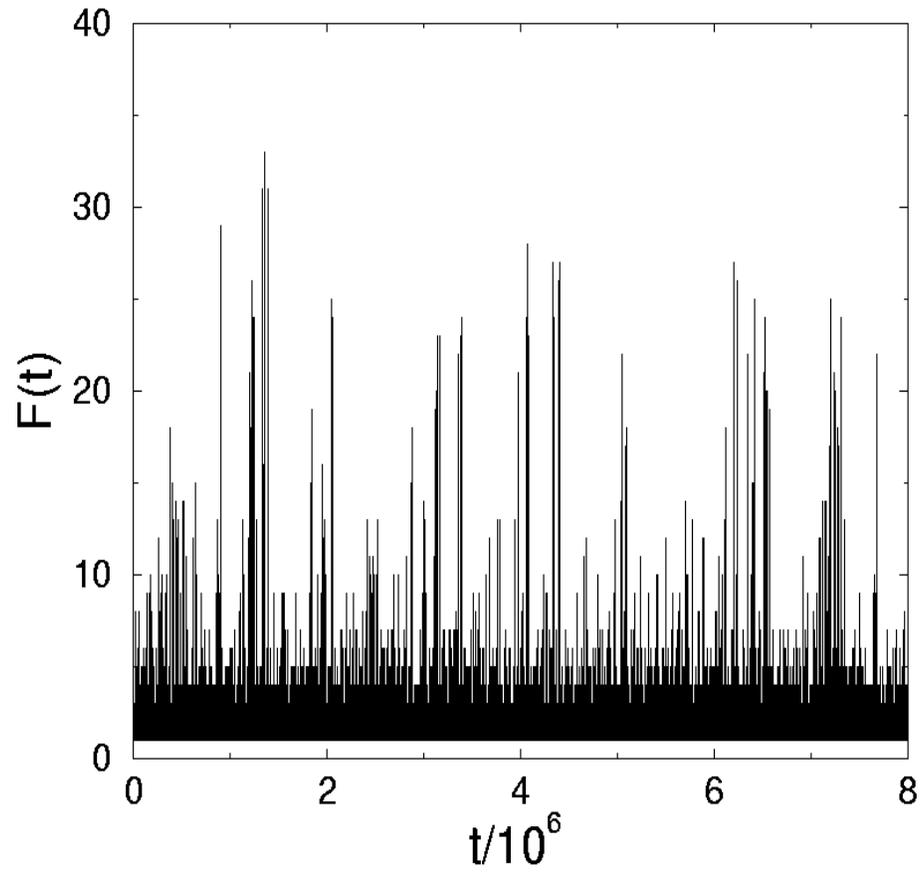}
%}
\caption{Time series of $F(t)$ for L=4096, $\Delta h=2^{-10}$ 
\label{F_t}}
\end{figure}

\begin{figure}[htb]
%\centerline
\includegraphics[width=12.0cm,height=12.0cm]{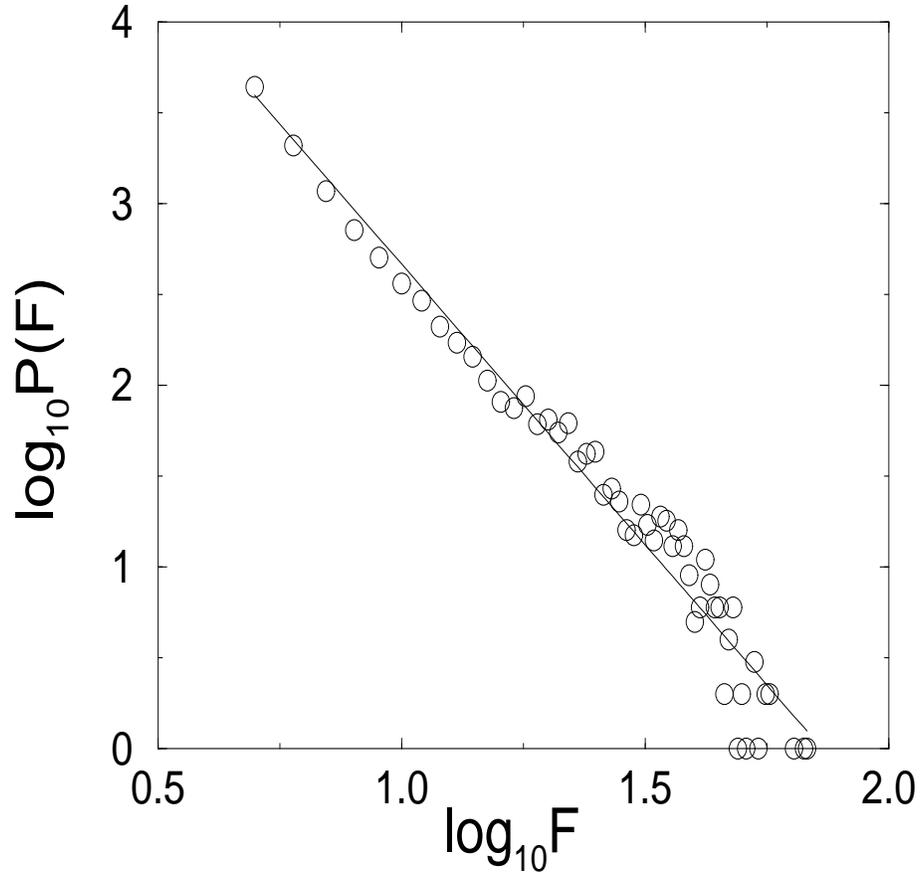}
%}
\caption{Distribution of forces $P_F(F)$ for L=8192, $\Delta h=2^{-10}$. 
The slope of the straight line fit is $-\mu=-3$.
\label{fP_F}}
\end{figure}

Finally we determine the time series of forces, defined as the number
of heights between $h_m(n)$ and  $h_m(n)-\Delta h$ as function
of time.  (Fig.~\ref{F_t}). 
The histogram of this time series is presented
in Fig.~\ref{fP_F} in a double logarithmic scale.
The slope of this plot is $\mu=3.0$, which is consistent with the theoretical 
prediction (\ref{mu}). 

\begin{figure}[htb]
%\centerline
\includegraphics[width=12.0cm,height=12.0cm,angle=270]{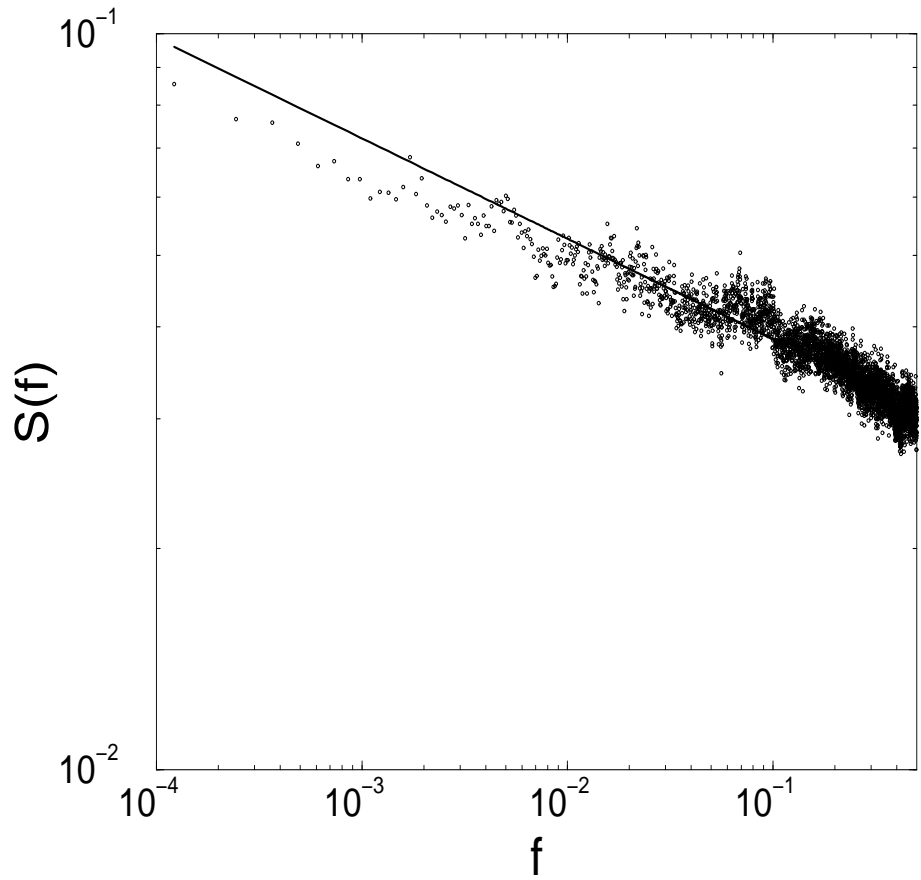}
%}
\caption{Power spectrum of the time series F(t) presented in Fig.(\ref{F_t}).
The slope of the straight line fit is $-\alpha -0.14$. 
\label{power}}
\end{figure}

Note that the time series $F(n)$ is slightly correlated, which can be
demonstrated by the negative slope of its power spectrum $S_F(f)\sim
f^{-\alpha}$ in the log-log scale (Fig.\ref{power}).  The explanation of this
phenomenon is based on the fact that values $h_m(n)$ fluctuate
 in the vicinity of
$h_c$ in a non-trivial way, so that $h_m(n)$ become less than 
$h_c+\epsilon$ at
time steps $n$ separated by intervals distributed according to
Eq. (\ref{P_s}). This is because these intervals coincide with
avalanches for the threshold $h_0=h_c+\epsilon$. 
The values of $h_m(n)$ below $h_c+\epsilon$
correspond to the large values forces $F(n)$ and thus the intervals between the
forces $F(n)$ above certain threshold 
are also distributed according to Eq. (\ref{P_s}). It can be shown
\cite{Paczuski,Lowen}that the exponent $\alpha$ of a time series generated by
peaks separated by intervals of zero signal distributed according to a
power law as in Eq. (\ref{P_s}) is equal to $\tau_s-1$ for $1<\tau_s<2$.
Thus according to Eq. (\ref{tau_s}) $\alpha=(d-1/\nu)/D\approx 0.13$.
Indeed, the numerical data of Fig.~\ref{power} give $\alpha\approx 0.14$ in a
very good agreement to the above theoretical prediction.  However, this value
of spectral exponent is much smaller than the values observed experimentally
which are in the range between 1 and 2.6.

This difference is to be expected since the materials in contact as well as
the strain gauge have finite elastic constants and inertia which produce a
time delay between the applied force and the displacement record by the
tribometer and lead to an effective integration of the input force time
series.  We would expect that if the materials were infinitely stiff then the
experimental force power spectrum should agree with the theoretical
predictions.  Therefore, we construct a mechanical model of a tribometer, 
that accounts for these effects.

We assume that the pin of the tribometer contacts the sample at time $t$ at a
point with coordinate $x(t)$ and it is dragged along the sample by the strain
gauge spring with spring constant $k$ attached to the body of the instrument
moving along the sample with constant velocity $v_0$, which is equivalent to
the rotational speed of the disk. The force measured by the tribometer is
thus $k[v_0 t-x(t)]$, which fluctuates as the pin moves against the sample with
velocity $v(t)=dx/dt$ and acceleration $a(t)=d^2x/dt^2$.  The equation of
motion of the pin is thus
\begin{equation}
m a= (v_0 t-x)k -F(t,v),
\label{ma}
\end{equation} 
where $F(t,v)$ is the friction force generated by the highest
asperity of the sample and $m$ is the mass of the pin.

Now our goal is to relate $F(t,v$) with the input from the Robin Hood
model. Note, that the physical time $t$ is not directly proportional to the
time step $n$ of the Robin Hood model, but is equal to the sum of the
durations of each time step
\begin{equation}
t=\sum_{i=1}^n t_i,
\label{time}
\end{equation}
where the durations $t_i$ are the times needed for the pin to travel a
characteristic distance $\Delta x$, which is the linear size of each
asperity.  We assume that if the pin moves along the sample by $\Delta x$ ,
the current asperity is destroyed and the landscape of the contact between
the pin and the sample is rearranged according to the rules of the
model. Thus the time step $n_t$ of the Robin Hood model, corresponding to a
given moment of time $t$ can be determined as $n_t={\rm int}[x(t)/\Delta x]$,
where ${\rm int}[...]$ denotes the integer part of the expression in the
brackets.

If $v=0$, $F(t,v)={\rm sign}(v_0 t -x)\min [b F(n_t),k(v_0 t -x)]$, where
$F(n_t)$ is the input from the Robin Hood model, and $b$ is some material and
load-dependent constant.  If $v\ne 0$, $F(x,v)= {\rm sign}(v) [b F(n_t)+ \eta
v]$, where $\eta>0$ is some dissipative constant.  Constant $b$ is
proportional to the load and depends on the elastic properties of the
material.  Introducing dimensionless variables by $x'=x/\Delta x$ and
$t'=tv_0/\Delta x$, we arrive to a dimensionless equation
\begin{equation}
a'=(t'-x')k'-F'(t',v'),
\label{ma'} 
\end{equation}
where $k'=k \Delta x^2/m v_0^2$ and $F'(t',v')$ is the same as $F(t,v)$ but
the constants $b$ and $\eta$ are changed by $b'=b\Delta x/m v_0^2$, $\eta'=\eta
\Delta x/mv_0$.  

\begin{figure}[htb]
%\centerline
\includegraphics[width=12.0cm,height=12.0cm,angle=0]{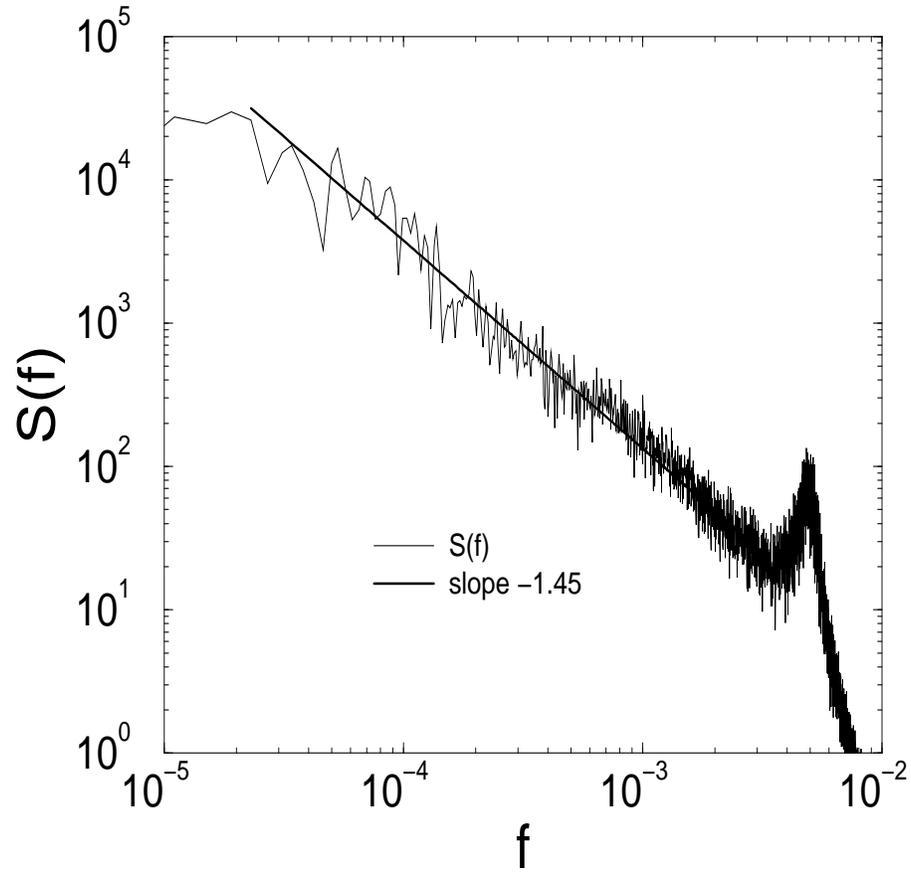}
%}
\caption{Power spectrum of the time series produced by Eq.(\ref{ma'}) for
$k'=0.001$, $b'=0.3$ end $\eta'=0.01$. 
\label{Sf}}
\end{figure}

Thus, there are three independent dimensionless parameters
of the model: $k'$, $b'$ and $\eta'$.  Varying these parameters, we found a
wide region in the parameter space in which the power spectrum of the model
resembles the experimental one.  A typical example of the spectrum for
$k'=0.001$, $b'=0.3$ and $\eta'=0.01$ is shown on Fig.(\ref{Sf}). The
frequency of the resonance peak is determined by $\sqrt{k'}/2\pi \approx
5\cdot 10^{-3}$. The peak becomes more pronounced as we decrease $\eta'$. The
increase in $\eta'$ also increases the slope of the spectrum. The increase of
$b'$ at given $k'$ increases the frequency region in which the power spectrum
follows the power law but it also increases the absolute value of the slope
closer to $2$, a characteristic value of the Brownian motion.  In general, an
integration of the time series corresponds to the increase of the spectral
exponent by 2, so the integration of the white noise produces the Brownian
noise. The observed spectral exponent $\alpha=1.45$ suggests that in a
certain range of parameters, our model acts as the fractional integrator of
the input signal.  For a very stiff spring and large dissipation ($k=1$,
$b=0.1$, $\eta=1$) the output signal of our equation is not much different
from the input time series $F(n_t)$ and we recover the small value of the
spectral exponent $\alpha=0.14$.

\section{Conclusions}  

We present experimental results and theoretical arguments that support the
presence of self organized criticality in dry sliding friction.  The
experiments are pin-on-disk friction force traces of aluminum-aluminum and
steel-steel systems.  In both cases and for a variety of normal loads, the
distribution of the friction force jumps and the frequency power spectra are
power laws.  The theoretical arguments are based on the application of the
Robin Hood model to the friction problem.  This model provides rules by which
the surface profile changes as a function of time.  The model introduces a
height $h$ that we interpret physically as the height of the asperity.  At
each time step, atoms from the highest asperity are distributed among
neighboring sites. We use the known distribution of heights and of maximal
heights $h_m$ of the Robin Hood model to obtain the time series of the
friction forces created by the asperities.  As the maximum height fluctuates
near the critical value, the number of smaller asperities whose heights are
within the reach of atomic forces also fluctuates, diverging as $h_m$ comes
close to the critical value $h_c$.  These smaller asperities correspond to
the contact sites and are responsible for the friction force.  Specifically,
the friction force is proportional to the number of contacts.  Thus we
propose that the friction force at a given time step is proportional to the
probability density of the interface heights at the current value of the
maximal height. The statistical distribution of the friction forces is
studied both numerically and analytically.

We also find that the large forces are bunched in time.  This is due to
fluctuations of the maximal heights above a constant critical
height.  When the maximum heights return to the critical value, the forces
become large.  Thus, the surface waxes and wanes between a situation of large
force due to many asperities acting, and a situation of smaller force in
which only few asperities are in contact.
  
In addition, we use the time series of 
forces as an input to the Newton's equation which describes 
the kinematics of the
pin.  For stiff or massless materials, the experimental distribution
of force jumps should coincide with the theoretical distribution of forces.
However, materials have finite mass and elasticity and thus the experimentally 
measured friction forces differ from the actual
forces at the contact. To investigate these effects, we
solve this equation numerically for different values of the parameters 
and find good agreement with the experiment.  

\section{Acknowledgments}.  

SVB thanks Yeshiva University for providing the high performance computer
cluster that made this work possible.  FRZ acknowledges support by Research
Corporation through grant CC5786.  FRZ thanks Phillip Abel, 
Mark Jansen, Kathleen Scanlon of NASA-Glenn Tribology Group for
collaboration in the initial stages of this project.

\eject

\begin{table}[htb]
\caption{\label{lattice} The values of exponents $\mu$ characterizing the
power law behavior of the distribution of the force jump sizes and spectral
exponents $\alpha$ characterizing the power spectrum of the friction force
time series for different materials and loads.}
\begin{ruledtabular}
\begin{tabular}{llll}
Material & Load & $\mu$ & $\alpha$ \\
\hline
M50  & 1000g. & 3.5  & 3.0  \\
Al   & 250g.  & 5.4  & 1.0  \\
Al   & 750g.  & 2.2  & 1.5  \\
Al   & 1000g. & 3.2  & 1.3  \\
\end{tabular}
\end{ruledtabular}
\end{table}

\end{document}